\documentclass[showpacs,aps,twocolumn]{revtex4}
\usepackage{bm}
\usepackage{amsmath}
\usepackage{graphicx}
\usepackage{subfigure}

\usepackage[usenames,dvipsnames]{color}
\definecolor{darkblue}{RGB}{0,0,196}
\usepackage[colorlinks=true,linkcolor=darkblue,citecolor=darkblue,urlcolor=darkblue]{hyperref}
\usepackage{setspace}
\usepackage{hyperref}
\usepackage{footmisc}
\usepackage[makeroom]{cancel}
\usepackage{comment}
\usepackage{lineno}
\usepackage{graphicx}
\usepackage{amsmath,bbm}
\usepackage{amssymb,bm}
\usepackage{yfonts}
\usepackage{lipsum}

\def\be{\begin{equation}}
\def\ee{\end{equation}}
\def\ba{\begin{eqnarray}}
\def\ea{\end{eqnarray}}

\newcommand\p{{\bm{p}}}

\newcommand \snn{$\sqrt{s}~$}
\newcommand  \pt {$\p_{\rm T}$}
\newcommand  \rt {$R_{\rm {T}}~$}
\newcommand  \rtt {$R_{\rm {T}}$}

\begin{document}

\title{Evolution of strange and multi-strange hadron production with \\ relative  transverse  multiplicity activity in underlying event}

\author{Prabhakar Palni\footnote{Email: Prabhakar.Palni@cern.ch, Prabhakar.Palni@unigoa.ac.in}$^{a,b}$}
\author{Arvind Khuntia\footnote{Email: Arvind.Khuntia@cern.ch}$^c$}
\author{Paolo  Bartalini\footnote{Email: Paolo.Bartalini@cern.ch}$^d$}
\affiliation{$^{a}$AGH University of Science and Technology, Faculty of Physics and Applied Computer Science, al. Mickiewicza 30, 30-059 Krakow, Poland}
\affiliation{$^b$Department of Physics, Goa University, Taleigao Plateau, Goa 403206, India}
\affiliation{$^c$H.~Niewodniczanski Institute of Nuclear Physics, Polish Academy of Sciences, 31-342 Krakow, Poland}
\affiliation{$^d$Instituto de Ciencias Nucleares, Universidad Nacional Autonoma de Mexico, Apartado Postal 70-543, Mexico D.F. 04510, Mexico}

\begin{abstract}

\noindent 
In this work, the relative Underlying Event (UE) transverse multiplicity activity classifier ($R_{\rm {T}}$)  is used to study the strange and multi-strange hadron production in proton-proton collisions. Our study with $R_{\rm {T}}$   would allow to disentangle these particles, which are originating from the soft and hard QCD processes. 
We have used the PYTHIA 8 Monte-Carlo (MC) with a different implementation of color reconnection  and rope hadronization models to demonstrate the proton-proton collisions data at  \snn = 13 TeV. The relative production of strange and multi-strange hadrons are discussed extensively in low and high transverse activity regions. In this contribution, the relative strange hadron production is enhanced with increasing \rtt. This enhancement is significant for the strange baryons as compared to mesons. In addition, the particle ratios as a function of \rt confirm the baryon enhancement in new Color Reconnection (newCR), whereas the Rope model confirms the baryon enhancement only with strange quark content. Experimental confirmation of such results will provide more insight into the soft physics in the transverse region, which will be useful to investigate various tunes based on hadronization and color reconnection schemes. 

\end{abstract}
\date{\today}
\maketitle

\section{Introduction}
\label{RT:intro}

In recent times, the most important discoveries in  proton-proton (pp) collisions are the evidence of collectivity \cite{Khachatryan:2016txc,alice2,Dumitru:2010iy,Ma:2014pva,Ortiz:2013yxa} and strangeness enhancement \cite{ALICE:2017jyt,alice6,Khuntia:2020aan}. They are remarkably similar to those observed in heavy-ion collisions at RHIC and LHC,  where these features are attributed to the production of a deconfined hot and dense medium, known as the Quark-Gluon Plasma (QGP) \cite{qgp1,qgp2}. The  measurement of strange hadron production in minimum-bias pp collision at the LHC energies has shown an enhancement as a function of multiplicity density \cite{Acharya:2018orn,ALICE:2017jyt,alice5,cms2,Aamodt:2011zza}. One of the important conclusions from that study was that most of the Quantum Chromodynamics (QCD) based MC models could not reproduce the data qualitatively.  Strange quarks in pp collisions can be produced in the hard scattering via flavor excitation/creation and gluon splitting \cite{Sollfrank:1995bn}. In soft scattering at low transverse momentum, strange quark pairs can be produced via non-perturbative processes. However, the production of strange quark being heavier is suppressed relative to hadrons containing only up and down quarks. The hadronic final states in minimum-bias events are the product of hard (perturbative) parton-parton scattering  and soft scattering which is called the Underlying Event (UE) \cite{rick,cdf3,cdf4}.  The major contribution to the UE activity comes from the initial and final state radiation,  beam-beam remnants, and PYTHIA specific physics processes such as multiple-parton interactions (MPI) in the same pp collision \cite{Sjostrand:MPI,paolo} and color reconnection (CR) \cite{Sjostrand:2013cya}.

 MPI plays a significant role in the particle production at the LHC energies \cite{alice6, enterria,Jpsi:MPI}. Measurements related to hadron yields in the minimum-bias (inelastic) collisions are convoluted with the hard scattering and the soft UE activity  \cite{Cuautle:2014yda}. Most of the minimum-bias collisions at the LHC energies are soft, with a typical transverse momentum scale less than 2 GeV/$c$  \cite{alice9,atlas:MB,cms:MB,cms:MB2}.

In this paper, we attempt to disentangle the hard QCD-process and investigate the hadron production. In particular, strange and multi-strange particles in the transverse region dominated by the soft QCD process. The novel relative transverse multiplicity variable (\rtt) has an excellent discriminating power to probe collective effects in hadronization \cite{tim}.   Furthermore, the multiplicity distribution associated with the UE exhibits Koba-Nielsen-Olesen (KNO) scaling \cite{antonio}. The relative transverse multiplicity variable, \rtt,~ serves as a perfect tool to study the strange hadron production yield and transverse momentum spectra from low to high underlying event activity. The UE measurement with charged-particle jets showed dependence on the  charged-particle jet radius, possibly due to the selection bias \cite{atlas1}. However, in this work, the highest transverse momentum charged-particle is used as a leading-object to define UE regions. The kinematical variables and acceptances are selected to match with ALICE detector \cite{alice0} to compare results directly with the ALICE, which has an excellent particle identification capabilities. Neutral strange particle production in underlying event as a function of the leading \pt~was first reported by CMS \cite{cms1} in pp collision~at $\sqrt{s} = $ 7 TeV. These results showed constant strange to charged-particle ratios beyond the plateau region.

In the current study, PYTHIA 8 (version 8.301) MC event generator along with three different models that incorporate various color flow and hadronization mechanisms to qualitatively describe various aspects of the underlying event activity is used. The  Monash 2013 tune (labeled as ``Monash")  \cite{monash} whose parameters are tuned to the latest LHC data (run I and II) is used as a reference model. In addition to particle decays and soft-QCD modeling, PYTHIA 8 features leading-logarithmic initial- and final-state parton showers,  Lund string hadronization, and MPI models \cite{Sjostrand:2007gs,pythia8html}. Furthermore, new QCD based color-space model, which allows strings to form between leading and non-leading color connected parton ~\cite{monash_newCR} is considered. When this model (which allows color-epsilon and anti-epsilon structure formation) is combined with PYTHIA’s model for junction fragmentation \cite{monash_newCR2}, it gives rise to a new source of baryon production. This novel model  (labeled  as  ``newCR") can describe the average transverse momentum of charged-particles and also production rates of baryons. The third model (labeled as ``Rope")  is based on the rope hadronization formalism~\cite{rope1,rope2}, which describes the interactions between the overlapping Lund strings in the transverse space to combine into color ropes. Such color ropes are expected to give more strange particles and baryons. In  these models, the minimum-bias (inelastic) events are generated for proton-proton collision at the center of mass energy, $\sqrt{s} =$ 13 TeV. 

The outline of this paper is the following. In Section II, we briefly describe the event selection criteria for UE and various observables under this study.  The results are discussed in Section III, based on the leading $p_{\rm {T}}$ and transverse multiplicity activity classifier, $R_{\rm {T}}$. Finally, in section IV, we summarize our findings and conclusions are drawn.

\section{Event selection and Observables}
The UE properties are derived based on the leading charged-particle direction in the event. This  charged-particle  is expected to have the direction of the parton produced with the highest transverse momentum in the hard scattering. The event  is classified into three topological regions: (i) Towards, (ii) Away and (iii) Transverse region in terms of the azimuthal angle difference, $\Delta \phi$, between the directions of the  leading charged-particle and that of any other particle in the event as shown in Figure~\ref{ue_cartoon}. The particles produced in the towards region are spanned by the azimuthal angle $|\Delta \phi| < 60^{\circ}$, and in the away region, $|\Delta \phi| > 120^{\circ}$, which is expected to be dominated by the hard scatterings. The UE activity can be best studied in the transverse region, $60^{\circ} < |\Delta \phi| < 120^{\circ}$. 
\label{ue_cartoon}
\begin{figure}[htp]
\begin{center}
	\includegraphics[width=85mm,scale=0.6]{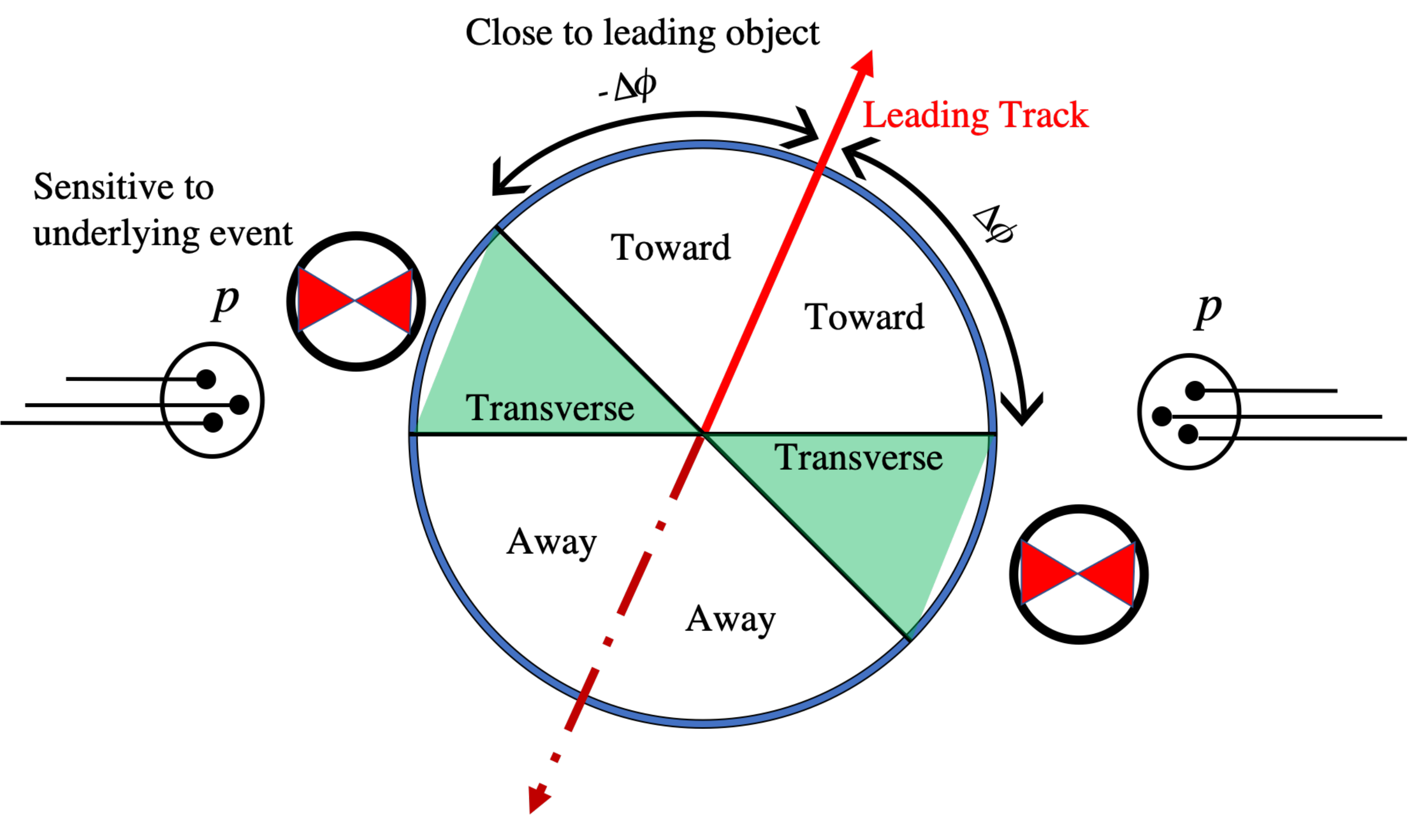}
	\caption{(Color online) Diagram illustrating the towards, transverse and away regions of the azimuthal angle with respect to the leading-charged particle in the typical pp collision event.}
	\label{ue_cartoon}
\end{center}
\end{figure}
\begin{figure}[!ht]
	\begin{center}
		\includegraphics[width=95mm,scale=0.85]{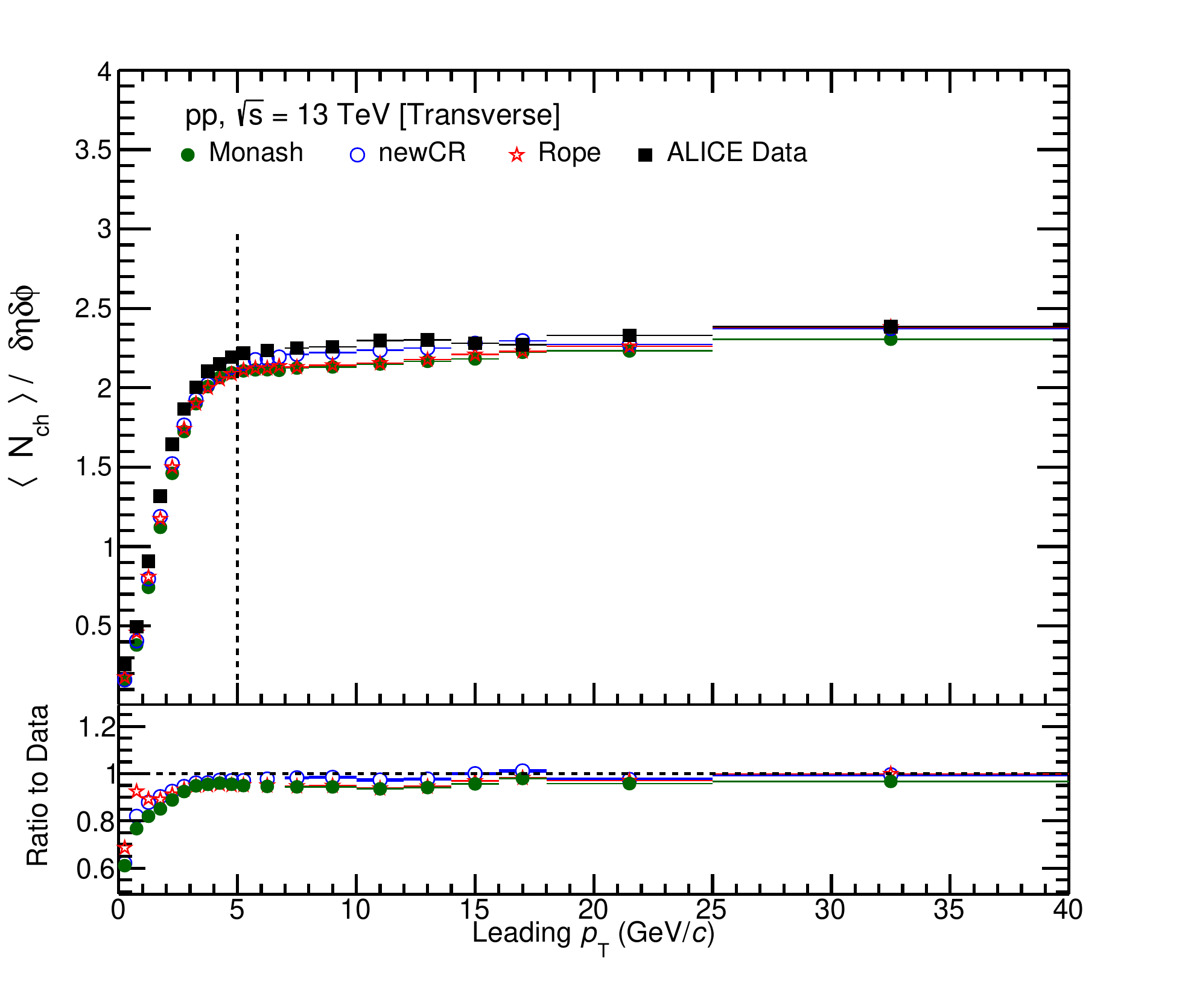}
		\caption{(Color online) The average number density of charged-particles as a function of  leading $\p_{\rm T}$ in the transverse region. The ALICE data are compared to three different models.}
		\label{strange:nch_comp}
	\end{center}
\end{figure}
\begin{figure*}[htbp]
\begin{center}
	\includegraphics[width=81mm,scale=0.3]{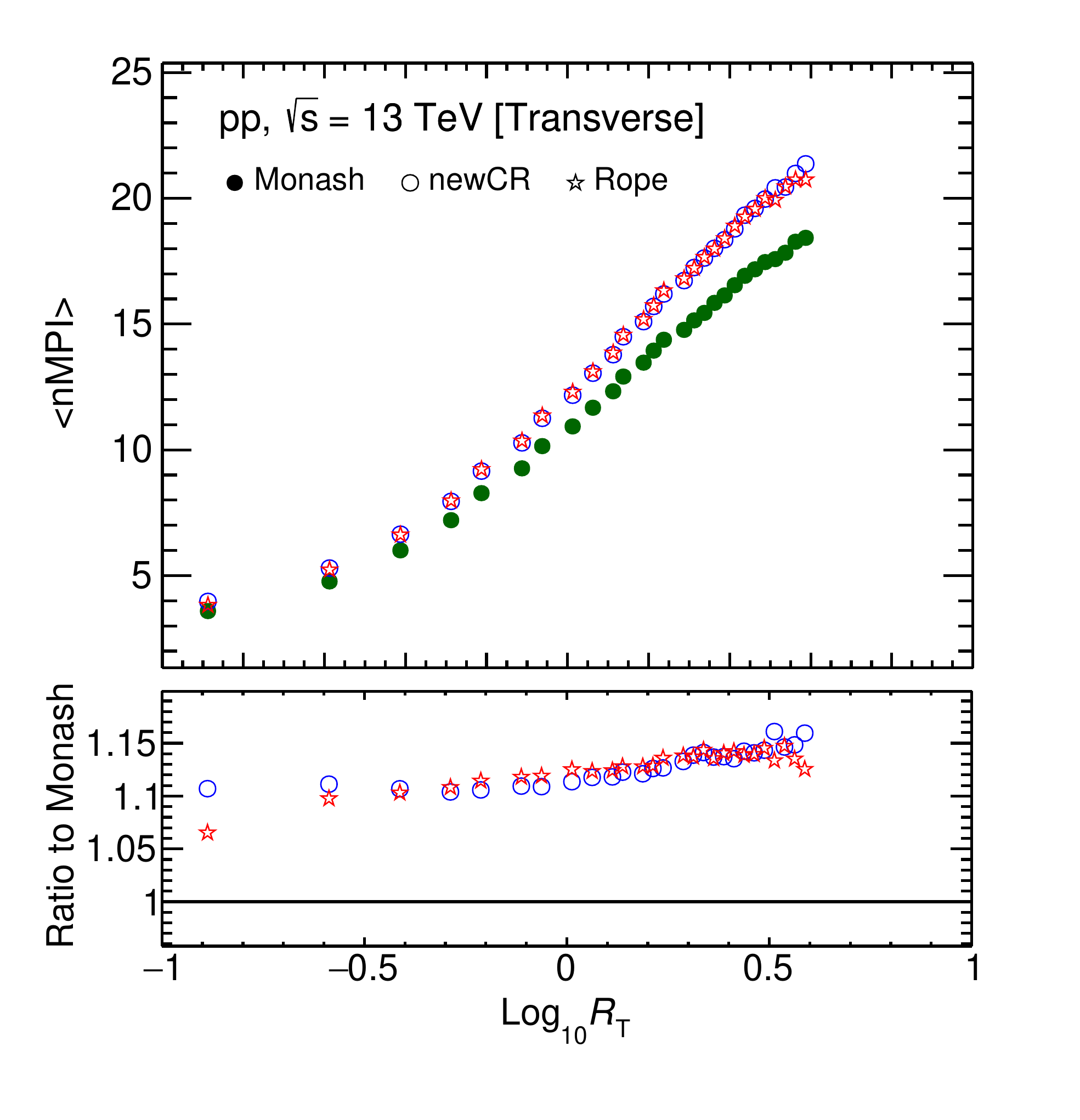}
	\includegraphics[width=80mm,scale=0.3]{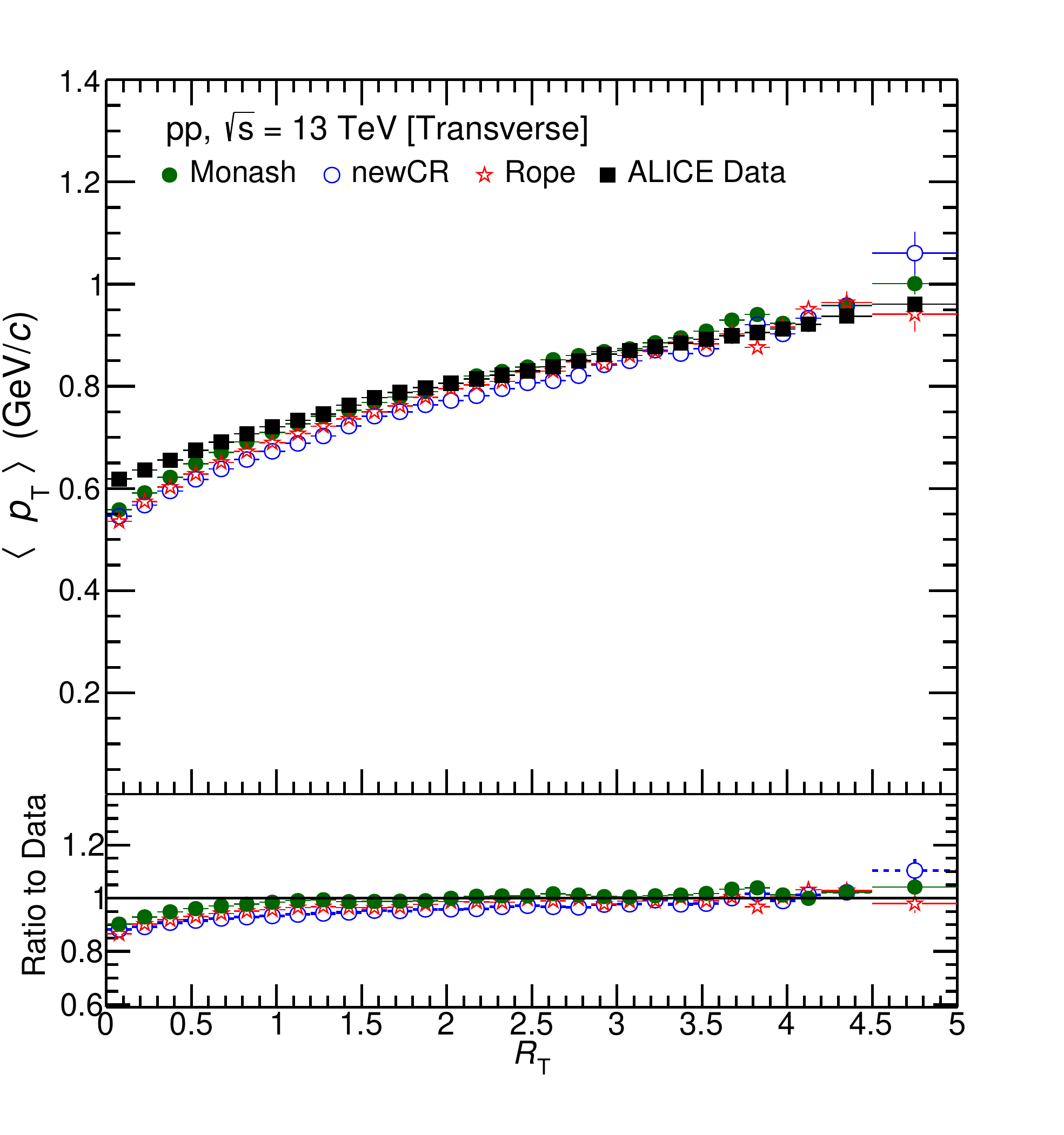}
	
	\caption{(Color online) The average number of MPI in an event as a function of $\log_{10}$(\rtt) for transverse region  (left panel). The $\langle$\pt$\rangle$ of charged-particles  as a function of \rt for the transverse region (right panel).}
	\label{rt:mpi}
\end{center}
\end{figure*}

All  the observables are measured using the primary charged-particle tracks   with transverse momentum \pt~ $>$ 0.15 GeV/$c$  and absolute pseudo-rapidity,  $|\eta| < 0.8$.  The particles  produced from the decay of hadrons with lifetime between $ 30 < \tau < 3000$  ps  are considered  as secondary particles. This study includes prompt neutral and charged strange hadrons ($K_{S}^{0}, K^{*0},  \phi, \Lambda (\bar{\Lambda}), \Xi^{\pm}, \Omega^{\pm}$) with   \pt~ $>$ 0.15 GeV/$c$ and $|\eta| < 0.8$.  

Traditionally, observables such as average multiplicity of charged-particles or average scalar sum of charged-particles \pt~per event are measured as a function of leading charged-particle track transverse momentum ($p_{\rm T}^{lead}$). 
The charged-particle density  in the transverse region rises steeply for low  values of $p_{\rm {T}}^{lead}$  and reaches a plateau ~\cite {atlas:ue,cms:ue,alice:UE}.
We have compared the charged-particle density measured by ALICE~\cite{alice:UE} in the transverse region to different PYTHIA 8 models as shown in the top panel of Figure~\ref{strange:nch_comp}. Agreement between the  ALICE data and the models are shown in the bottom panel.  For $p_{\rm {T}}^{lead}$ below 2 GeV/$c$, all the three models under-predict the ALICE data, whereas  in the plateau region differences between the data and models is below $\sim$ 8\% and independent of  $p_{\rm {T}}^{lead}$. The UE activity is classified in the plateau region ($ 5 <  p_{\rm T}^{lead} < 40 $ GeV/$c$) to define relative transverse activity variable, \rtt. In this region, all the models are in good agreement with the experimental data. The \rt  is defined as the ratio of multiplicity of the inclusive charged-particles and identified strange hadrons ($N_{inc}$) to its event-averaged multiplicity ($\langle N_{inc} \rangle $) in the transverse region:
\begin{eqnarray}
 R_{\rm T}= \frac{ N_{inc}}{ \langle N_{inc} \rangle}  
\end{eqnarray}
 %
 
  The \rt variable is a useful tool to differentiate events with higher-than-average  (\rt $> 1$) from lower-than-average (\rt $< 1$) UE activity, irrespective of the center of mass energy or any fiducial requirements on the observable. 
   The event-averaged multiplicity, $\langle N_{inc} \rangle$, (and width) for  ``Monash", ``newCR", and ``Rope" models are 7.72 (4.94), 7.97 (4.94), and 7.65 (4.85), respectively. For comparing ALICE experimental data with MC models (as shown in Fig. \ref{strange:nch_comp} and \ref{rt:mpi} (right panel)), we have considered the charged-particles instead of $N_{inc}$.
\section{Results and Discussion}
\label{RT:results2}

In this work, all the results are discussed by taking Monash as a reference model to understand the particle production mechanism with respect to newCR and Rope models. Figure~\ref{rt:mpi} (left panel) shows average number of  MPI in an event as a function of $\log_{10}$(\rtt)~for the transverse region in three different models. The average number of MPI increases steeply with $\log_{10}$(\rtt) beyond 0,   which corresponds to \rt $\sim 1$ and it is 10-15\% higher for newCR and Rope models. The highest event activity at $\log_{10}$(\rtt) $\simeq$ 0.5 has twice the average number of MPI  as compared  to $\log_{10}$(\rtt) = 0. The mean transverse momentum ($\langle$\pt$\rangle$)~of the charged-particles  as a function of $R_{\rm {T}}$ in the transverse region is compared among different MC models   as shown in Figure~\ref{rt:mpi} (right panel).  The $\langle$\pt$\rangle$ increases with $R_{\rm {T}}$ for all the three MC models and qualitatively explains the ALICE data. The color reconnection present in all the three models allows interaction between the strings which generates flow-like effects in the final state. An increase in the $\langle$\pt$\rangle$ with $R_{\rm {T}}$ is attributed to the presence of CR between the interacting strings. The ratio of ALICE data to MC models is shown in the bottom panel of Figure~\ref{rt:mpi}.  The $\langle$\pt$\rangle$ is  underestimated by all the three models by less than 10\% below $R_{\rm {T}} < 0.5$, whereas for $R_{\rm {T}} >  3$,  the trend is in agreement with the ALICE data~\cite{alice:UE}.  A similar trend has been observed for the $\langle$\pt$\rangle$ as a function of charged-particle density in minimum bias events between the ALICE data \cite{alice:meanpt} and Monash model. 

\begin{figure}[htbp]
	\begin{center}
		\includegraphics[width=85mm,scale=0.90]{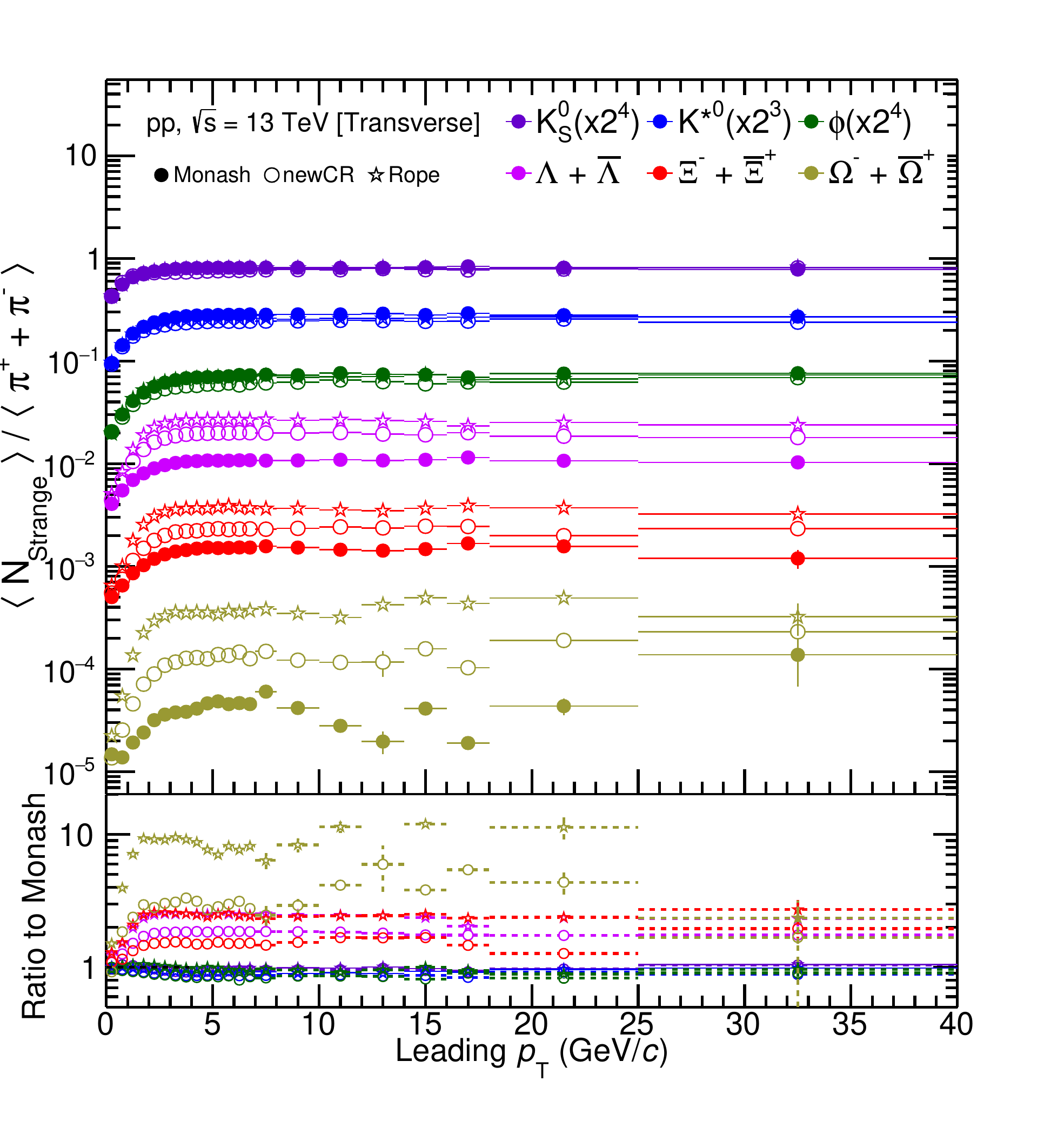}
		\caption{(Color online) Ratio of average strange hadrons to  pions as a function of  leading $\p_{\rm T}$ in the transverse region.}
		\label{strange:leadpt}
	\end{center}
\end{figure}

The average ratios of production rates of the strange hadron to the average rates of the pions as a function of $p_{\rm T}^{lead}$ in the transverse region are shown in Figure~\ref{strange:leadpt}. To probe any sensitivity to the relative strangeness enhancement, the strange hadron yields are normalized to pion yields. Moreover, this ratio also factors out any contribution from differences in the overall multiplicity spectra of the tunes and MC models. The average strange particle production relative to pions increases significantly at low lead \pt, eventually reaching the plateau region at around 1-3 GeV/$c$. The average ratios for different particles are scaled by different factors to improve visibility. The similarity between production rates of strange hadrons ratios and charged-particles \cite{alice:UE}  as a function of $p_{\rm T}^{lead}$ confirms the impact-parameter picture of the multiple parton interactions in pp collisions, in which the centrality of the collision and the MPI activity are correlated. Strange hadrons with higher mass attain the plateau at higher leading \pt. The newCR and Rope models predict higher production of strange baryons as compared to strange mesons. The production rates of  $\Omega$  ($|S|$ = 3) with $p_{\rm T}^{lead}$  in the plateau region for newCR and Rope models are $\sim$ 10  and $\sim$ 3 times higher as compared to Monash, respectively. The relative yields are similar in the case of  $\Lambda$ and $\Xi$ in the Rope model. However, higher production rates ($\sim$ 30\%) are predicted in the newCR model for $\Lambda$ as compared to $\Xi$.
\begin{figure*}[htbp]
\begin{center}
	\includegraphics[width=85mm,scale=0.50]{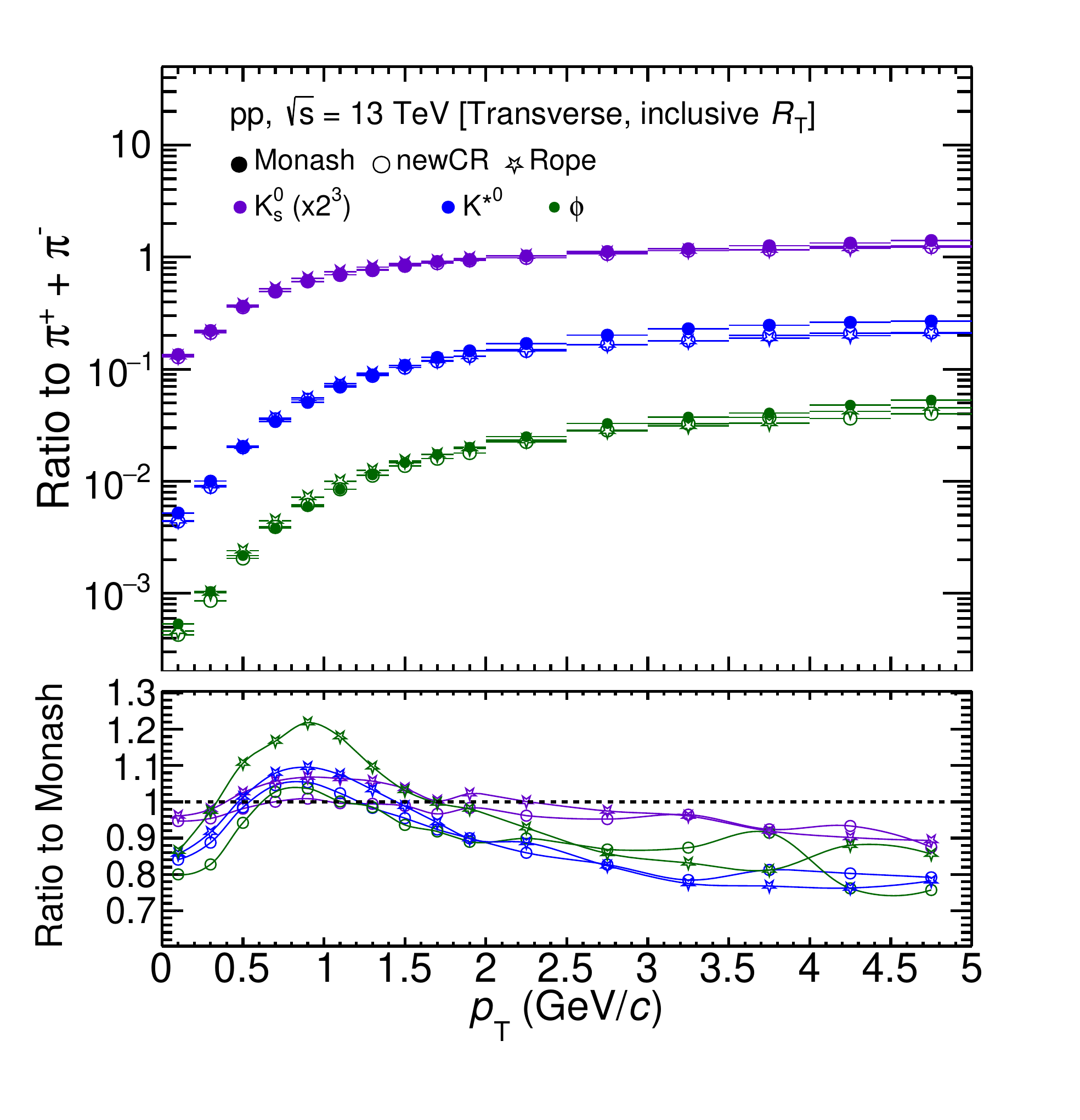}
	\includegraphics[width=85mm,scale=0.50]{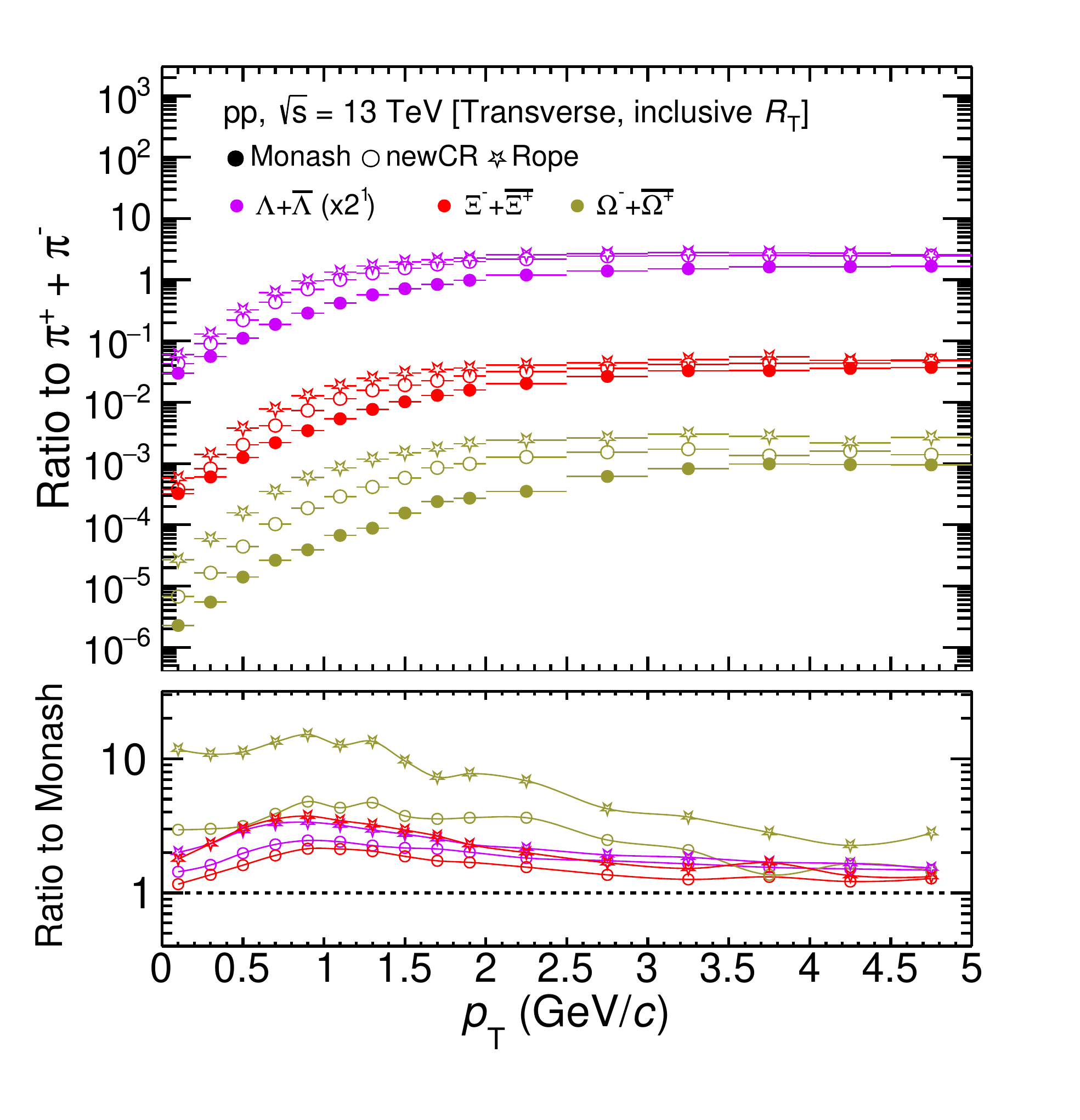}
	\caption{ (Color online) \pt-differential ratios of strange mesons (left panel) and strange baryons (right panel) to pions for inclusive \rt in the transverse region. The \pt-differential ratios for $K_{S}^{0}$ and $\Lambda$ are scaled by different factors to improve the visibility.}
	\label{pt:diff}
\end{center}
\end{figure*}
The particle production in different \pt~region has different origins and thus \pt-differential ratios provide a more differential understanding of particle production and hadronization. The \pt-differential ratio of the strange and multi-strange hadrons to pions in the transverse region for integrated \rt is shown in Figure \ref{pt:diff}. A universal trend in the relative production of strange hadrons with \pt~  can be seen, which finally reaches a plateau beyond 2-4 GeV/$c$ for all the three models.  The plateau shifts towards the high \pt~value for the heavier strange particles, which can be attributed to collectivity like effects originating from the color reconnection by boosting the emitted particles to higher momenta.
Comparison with the Monash shows an increase in the strange meson production by (10-20 \%) in \pt~ region 0.5-1.5 GeV/$c$ as shown in  Figure \ref{pt:diff} (left panel) for Rope model. However, in higher \pt~region production rates in both Rope and newCR models are suppressed. Significant differences in strange baryons rates are observed among the MC models as shown in Figure \ref{pt:diff} (right panel). The magnitude is highest for the $\Omega$, by up to a factor of 10 for the Rope,  and  4  for the newCR model in the   \pt~region between 0-3 GeV/$c$.  Comparison of the Rope and newCR models for $\Lambda$ and $\Xi$ show 2-3 times higher production rates in the \pt~region 0-3 GeV/$c$.
The behavior of the strange hadron production normalized to pions with \rt is a useful tool to understand different hadronization models. A comprehensive study has been performed for strange hadrons both in low (\rt $<$ 1)  and high \rt (\rt $>$  1)  regions which correspond to low and high UE activity in the transverse region.
\begin{figure*}[htbp]
\begin{center}
	\includegraphics[width=85mm,scale=0.50]{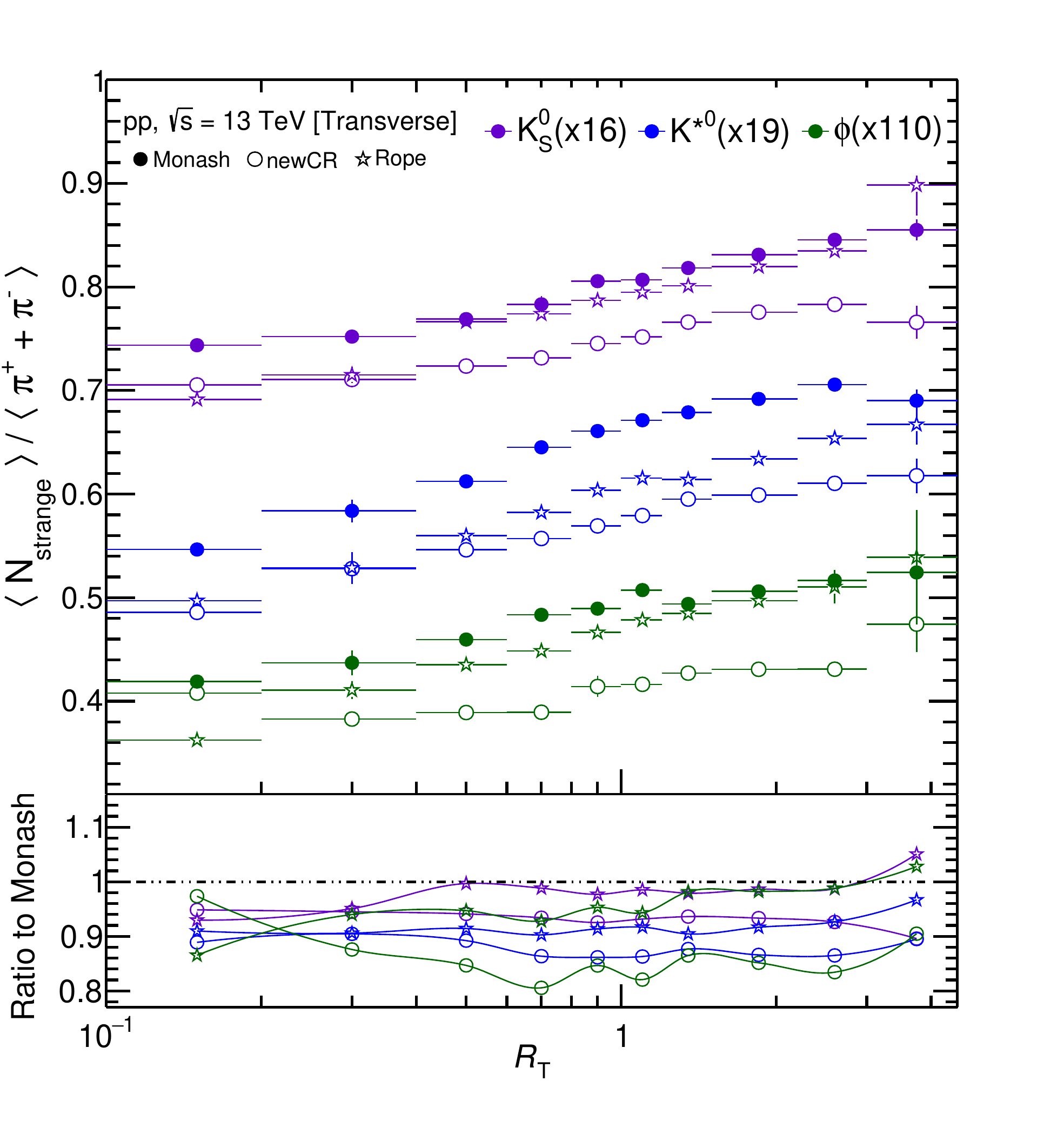}
	\includegraphics[width=85mm,scale=0.50]{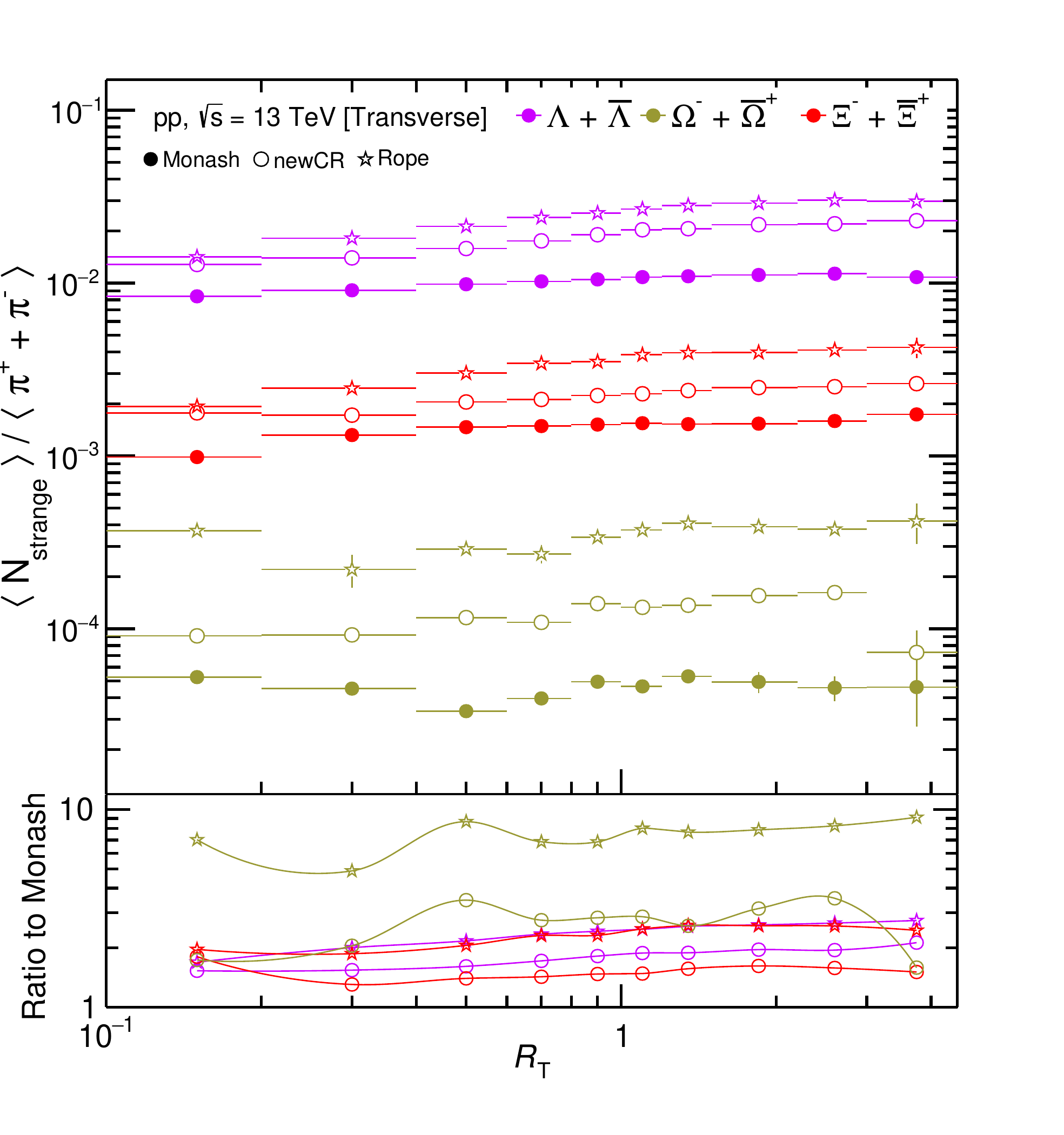}
	\caption{(Color online) Ratios of strange particles to pions as a function of $R_{\rm {T}}$ in the transverse region. The ratios of $K_{S}^{0}, K^{*0}$ and $\phi$  are scaled by different factors for better visibility.}
	\label{strange:rt}
\end{center}
\end{figure*}
The ratio of average strange hadron yield to pions as shown in Figure~\ref{strange:rt}, increases with increasing \rt for all the MC models. This increase in yield ratios with  \rtt, particularly for \rt $> 1$ can be attributed to the higher multiple partonic interactions as shown in Figure~\ref{rt:mpi}  (left panel). In the case of strange mesons, Monash shows higher production rates with \rtt, whereas Rope and newCR model shows 10-15\% less production. On the other hand, for strange baryons, enhancement is significantly higher as compared to mesons.  The highest production rate is predicted for multi-strange baryon ($\Omega$) in the Rope  followed by newCR model. This trend is consistent with our earlier observation with leading \pt, as shown in Figure~\ref{strange:leadpt}. Enhancement in newCR as compared to Monash can be understood as an introduction of the junctions structure during the color reconnection phases, which favors the baryon enhancement over mesons. Further contribution to strange and multi-strange hadrons in the Rope model comes from additional production of strangeness and di-quarks due to interacting strings which form ropes and consequently hadronize.
\begin{figure}[!ht]
\begin{center}
	\includegraphics[width=85mm,scale=0.6]{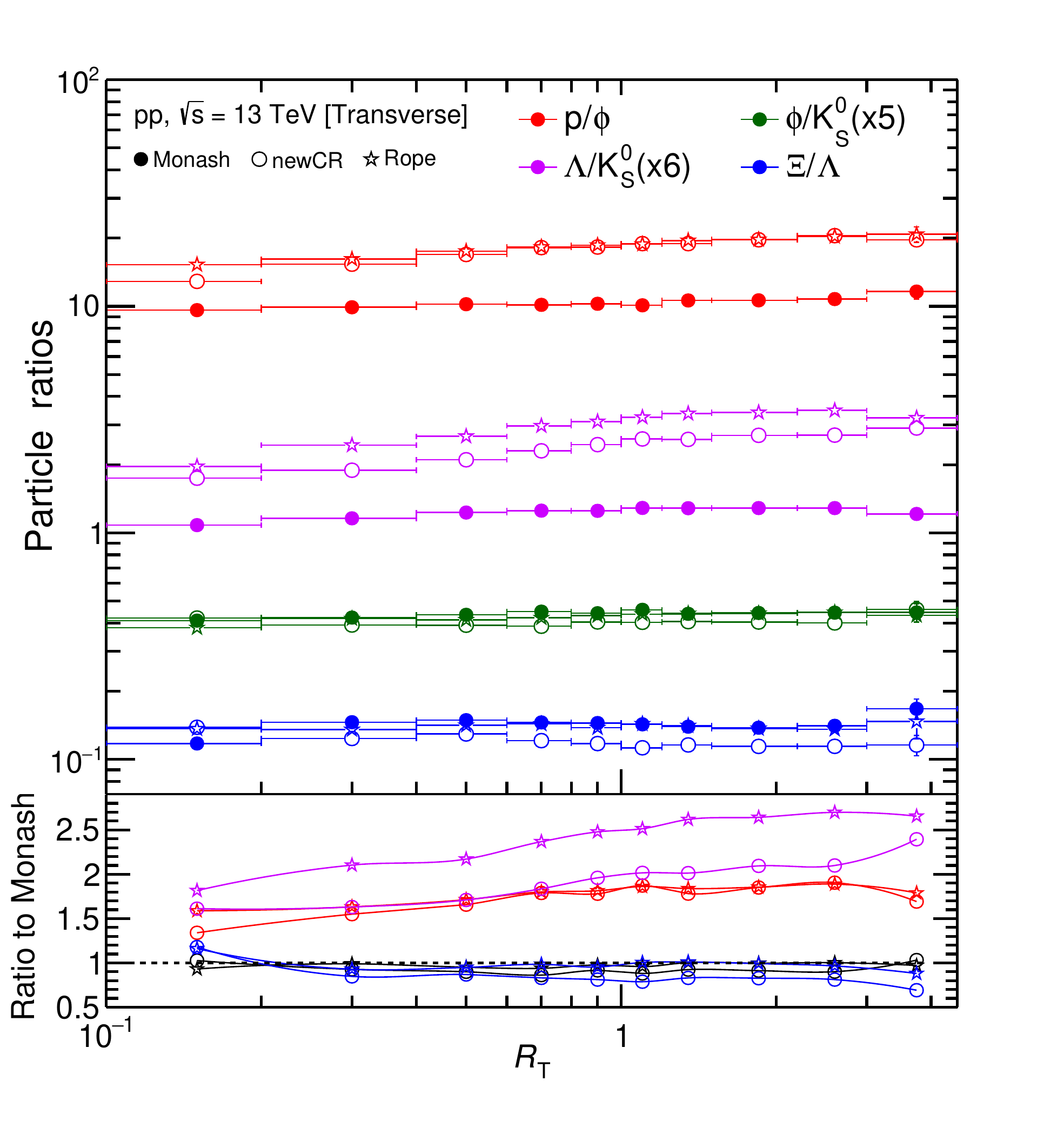}
	\caption{(Color online) An average hadron yield ratios as a function of $R_{\rm {T}}$  in the transverse region. The $\Lambda/{\rm{K_{S}^0}}$ and  $\phi/{\rm{K_{S}^0}}$  ratios are scaled by different factors to improve the visibility. }
	\label{hadron:rt}
\end{center}
\end{figure}
The particle production chemistry in UE is investigated by measuring different particle yield ratios which are sensitive to different hadron masses, baryon/meson, and strangeness content with \rtt, as shown in Figure~\ref{hadron:rt}. We observe  dependence for baryon to meson ratios  as a function of \rt for p/$\phi$ and $\Lambda/{\rm{K_{S}^0}}$.  The $\Lambda/{\rm{K_{S}^0}}$ ratio has the highest production rate in the Rope model ($\sim$ 2.5 times) followed by newCR ($\sim$ 2 times) as compared to Monash in the high relative transverse activity region. 
The p/$\phi$ ratio obtained from the newCR and Rope models has similar production rate as proton being non-strange particles do not have further contributions from the Rope model. The individual particle yield increases with \rtt, however, the effective baryon to baryon ($\Xi/\Lambda$)  or meson to meson ($\phi/{\rm{K_{S}^0}}$) ratios cancel-out, which results in the independent behavior with \rtt.
\section{Summary and Conclusion}
\label{RT:summary}
We have extensively studied the strange and multi-strange particle production in the transverse region as a function of leading \pt~and transverse multiplicity activity in pp collisions at $\sqrt{s}$ = 13 TeV. Three different models with different implementation of hadronization and color reconnection are used to study the dependence of the particle production in the transverse region using the PYTHIA 8 event generator. 
In this paper, an attempt has been made to disentangle the soft and hard QCD processes and to look into the particle production in a more differential manner using the transverse UE activity classifier, \rtt. 
The newCR model includes baryon enhancement due to the junctions structure introduction during the color reconnection stage. Whereas Rope model incorporates strangeness enhancement due to surplus production of strangeness and di-quarks compared to Monash as a function of \rtt. 
The steep rise of $\langle$\pt$\rangle$ in the transverse region as a function of \rt  is qualitatively explained by all the three models. These models underestimate the $\langle$\pt$\rangle$ at low transverse activity region (\rt $<1$)  for the ALICE data,  whereas it is in good agreement  at higher \rt values. The average strange particle production relative to pions as a function of leading \pt~ increases significantly at low leading  \pt, eventually reaching the plateau region at around 1-3 GeV/$c$. A systematic shift in the knee of the plateau towards the higher value of leading \pt ~with the heavier hadrons is observed. An enhancement of baryons relative to pions with leading \pt~ is predicted in the newCR model and additional production in the Rope model. The production rate of baryon with the highest strangeness, $\Omega$, in the plateau region for newCR and Rope models is $\sim$ 10  and $\sim$ 3 times higher as compared to Monash, respectively. The \pt-differential particle ratios in the transverse region show the evolution of the spectral shape for both strange mesons and baryons in low-\pt~region between 0-3 GeV/$c$. A strong increased production of strange baryons in the newCR and Rope model around 1 GeV/$c$ is observed. The ratio of average strange hadron yield to pions increases as a function of \rt in all the MC models.  The Monash shows higher production rates with \rt for strange mesons, whereas the Rope and newCR models show higher production rates for strange baryons. The p/$\phi$ and $\Lambda/{\rm{K_{S}^0}}$ ratios as a function of \rt confirms the baryon enhancement in newCR, whereas Rope model is responsible for baryon enhancement with strange quark content. Moreover, meson to meson and baryon to baryon ratios do not show a significant dependence on \rtt.

 Previous studies show that most of the MC models do not explain the enhancement of strange hadron production in pp collision as a function of charged-particle multiplicity density. Our study with the transverse multiplicity activity classifier provides qualitative evidence of an increase in strange hadron production in the underlying event. Experimental confirmation of these results will provide more insight into the soft physics in the transverse region as a function of \rtt. This will further help to tune various hadronization and color reconnection schemes which can better explain particle production in pp collision.
 \\
\section*{Acknowledgements}
The authors acknowledge the support from PL-Grid Infrastructure. Prabhakar Palni acknowledges the support in part by Polish National Science Centre grant DEC-2016/23/B/ST2/01409, by the AGH UST statutory tasks No. 11.11.220.01/4 within subsidy of the Ministry of Science and Higher Education. AK acknowledges the support in part by Polish National Science Center grant NCN HARMONIA, UMO-2016/22/M/ST2/00176 (ALICE), 2017-2020.


\end{document}